\documentstyle[12pt]{article}
\input mssymb
\begin{document}
\pagestyle{plain}
\pagenumbering{arabic}
\begin{center}
{\Large\bf Some Applications of Differential}\\
\vspace*{0.3cm}
{\Large\bf Topology in General Relativity}\\
\end{center}
\vspace*{0.4cm}
\begin{center}
{\large Andrew Chamblin}
\end{center}
\vspace*{0.1cm}
\begin{center}
{\it Department of Applied Mathematics and Theoretical Physics,}\\
{\it University of Cambridge, Cambridge CB3 9EW, England}\\
\end{center}
\vspace*{0.5cm}

\begin{center}
{\bf Abstract}\\
\end{center}

{Recently, there have been several applications of differential and
algebraic topology to problems concerned with the global structure of
spacetimes. In this paper, we derive obstructions to the existence of
spin-Lorentz and pin-Lorentz cobordisms and we show that for
compact spacetimes with non-empty boundary there is no relationship
between the homotopy type of the Lorentz metric and the causal
structure. We also point out that spin-Lorentz and tetrad cobordism are
equivalent.
Furthermore, because the original work [7] on metric homotopy and causality may
not
be known to a wide audience, we present an overview of the results here.}\\
\vspace*{0.6cm}

\begin{center}
{\bf Contents}\\
\end{center}
{\noindent {\bf I.}~~ Definitions and Conventions}\\
\\
{\noindent {\bf II.}~~ Equivalence of tetrad and spin-Lorentz cobordism}\\
\\
{\noindent {\bf III.}~~ Derivation of the obstruction to spin-Lorentz
cobordism}\\
\\
{\noindent {\bf IV.}~~ Discussion of Clifford algebras}\\
\\
{\noindent {\bf V.}~~ Derivation of the obstruction to pin-Lorentz cobordism}\\
\\
{\noindent {\bf VI.}~~ Applications of the obstructions}\\
\\
{\noindent {\bf VII.}~~ Kinking and causality}\\
\\
{\noindent {\bf VIII.}~~ Conclusion}\\
\\
{\noindent Acknowledgements}\\
\\
{\noindent References}\\
\vspace*{0.6cm}

\begin{center}
{\bf I. ~Definitions and Conventions}\\
\end{center}

{In this paper, by the word `spacetime' we shall mean a
four-dimensional manifold $M$, connected and smooth (though not necessarily
orientable), possessing an everywhere non-singular Lorentz metric $g_{ab}$.

The existence of an everywhere non-singular Lorentz metric on a time-orientable
$M$ is
equivalent to the existence of a global non-vanishing vector field
$v$. To see this, recall that the underlying Riemannian manifold $M$
possesses a Riemannian metric, $g^{R}_{ab}$. Given a vector field $v$,
one can define the Lorentz metric, $g_{ab}$, in terms of the
Riemannian metric and $v$ via the relation}
\begin{equation}
{g_{ab} = g^{R}_{ab} ~-~ {\frac{2v_{a}v_{b}}{(g^{R}_{ab}v^{a}v^{b})}}}
\end{equation}
{The converse follows by diagonalising the given Lorentz metric into
`Riemannian metric' and (negative eigenvalue) `eigenvector' parts, and
defining $v$ to be the vector with negative eigenvalue (see [1] or [14]).

We assume{\footnote{Note: It is not necessary (generally) to assume
time-orientability; we could still define a notion of kinking for
non-orientable $M$.}} (for the time being) that our spacetimes are
time-orientable,
i.e. that we can make a globally consistent choice for the sign of $v$ (one
cannot
propagate $v$ around some closed loop in $M$ and end up with $-v$).

Broadly speaking, the kink number is an integer which classifies
metrics up to homotopy. To make this more precise, let ($M, g_{ab}$)
be a spacetime and ${\Sigma} ~{\subset}~ M$ a three-dimensional,
connected, orientable submanifold. Since ${\Sigma}$ three-dimensional
and oriented, we can always find a global framing {\{}$u_{i}: ~i = 1,
2, 3${\}} of ${\Sigma}$ together with a unit normal, $n$, to
${\Sigma}$. We can then extend this tetrad framing ($n, u_{i}$) of
${\Sigma}$ to a collar neighbourhood}
\[
{N ~{\cong}~ {\Sigma} ~{\times}~ [0, 1].}
\]
{(We extend to $N$ to deal with the case ${\Sigma} ~{\cong}~
{\partial}M$). Let $v$ be the unit timelike vector determined by
$g_{ab}$; then $v$ can be written}
\begin{equation}
{v = v^{0}n ~+~ v^{i}u_{i}}
\end{equation}
{such that ${\displaystyle{\sum_{i}}}{(v^{i})^{2}} = 1$. Clearly, then, $v$
determines
a map $K: ~{\Sigma}  ~{\longrightarrow}~ S^{3}$, by assigning to each point $p
{}~{\in}~
{\Sigma}$ the direction in $T_{p}M$ (a point on the $S^{3}$ determined by the
tetrad
$(n, u_{i})$) that $v_{p}$ points to, i.e., visually: Fig. 1}\\
\\
{\noindent The north pole of $S^{3}$ is given by $n$. We then define the kink
number
of $g_{ab}$ with respect to ${\Sigma}$ as}
\begin{equation}
{{\mbox{kink}}({\Sigma}; g_{ab}) = {\deg}(K)}
\end{equation}
{where deg$(K)$ is `the degree of the mapping $K$'.}\\
\\
{{\bf Convention}~ {\it If $v$ is a timelike vector determined by $g_{ab}$,
we shall often write}}
\[
{{\mbox{\it kink}}({\Sigma}; g_{ab}) = {\mbox{\it kink}}({\Sigma}; v).}
\]
\\
{Now, for our immediate purposes we shall be concerned with kinking with
respect to
${\partial}M$, the boundary of our spacetime. In particular, we shall be
concerned
with the case $M$ compact, with}
\begin{equation}
{{\partial}M ~{\cong}~ {\Sigma}_{0} ~{\cup}~ {\Sigma}_{1} ~{\cup}~ ... ~{\cup}~
{\Sigma}_{n}}
\end{equation}
{where the ${\Sigma}_{i}$'s are now closed, connected, oriented three-manifolds
and
`${\cup}$' is the operation of disjoint union. We wish to define the quantity
kink$({\partial}M; g_{ab})$. On differential topological grounds (see [4]) we
see that
it makes sense to write}
\begin{equation}
{{\mbox{kink}}({\partial}M; g_{ab}) = {\sum_{i}}~{\mbox{kink}}({\Sigma}_{i};
g_{ab})}
\end{equation}
{{\it once} we have decided on some convention for choosing the sign of
$n_{i}$ (the unit normal to each ${\Sigma}_{i}$) consistently. Our convention
is
simply that $n_{i}$ is always pointing {\it out} of $M$. Having established
this, we can now discuss the concept of cobordism. First, however, we recall
the
following}\\
\\
{\noindent {\bf Definition}~ {\it Let $M$ be an oriented and time-orientable
spacetime, with orthonormal frame bundle $O(M)$ a principal bundle with
structure group
$SO(1, 3)_{0}$. We say that $M$ has {\underline{$SL(2, {\Bbb C})$ - spin
structure}} iff there
exists a principal bundle ${\bar O}(M)$ (with structure group Spin$(1, 3)_{0}
{}~{\simeq}~ SL(2, {\Bbb C})$) which is a $2 - 1$ covering of $O(M)$, so that
the following
diagram commutes:}}\\
\[
\begin{array}{cccccc}
SL(2, {\Bbb C}) ~{\simeq}~ {\mbox{\it Spin}}(1, 3)_{0} &{\longrightarrow}
&{\bar O}(M)
&{\longrightarrow} &M & \\
{\downarrow} & &{\downarrow} & &{\downarrow} &{\mbox{\it identity}} \\
SO(1, 3)_{0} &{\longrightarrow} &O(M) &{\longrightarrow} &M &
\end{array}
\]
\\
{\noindent {\it We shall call such an $M$ a {\underline{spin-Lorentz
manifold}}.

Now, suppose we are given a collection of three-dimensional, connected,
orientable,
closed manifolds ${\Sigma}_{1}, {\Sigma}_{2}, ... {\Sigma}_{n}$. We say that
there is
a {\underline{spin-Lorentz}} {\underline{cobordism}} for {\{}${\Sigma}_{i}: ~i
= 1, ...
n${\}} iff there exists a spin-Lorentz manifold $M$ satisfying}}
\[
{{\partial}M ~{\cong}~ {\Sigma}_{1} ~{\cup}~ {\Sigma}_{2} ~{\cup}~ ... ~{\cup}~
{\Sigma}_{n}}
\]
\\
{We have the related}\\
\\
{\noindent {\bf Definition}~ {\it Let $M$ be a smooth, four-dimensional
Riemannian manifold. We say that $M$ is {\underline{parallelizable}} iff there
exists a
global non-vanishing tetrad field, {\{}$e_{i}${\}}, on $M$.

Let ${\Sigma}_{1}, ... {\Sigma}_{n}$ be a collection of three-manifolds as
above. We
say that there is a {\underline{tetrad cobordism}} for {\{}${\Sigma}_{i}: ~i =
1, ...
n${\}} iff there exists a parallelizable four-manifold $M$ such that}}
\[
{{\partial}M ~{\cong}~ {\Sigma}_{1} ~{\cup}~ {\Sigma}_{2} ~{\cup}~ ... ~{\cup}~
{\Sigma}_{n}.}
\]
\\
{\indent More generally, we can consider the problem of finding cobordisms
admitting
other types of structures; the study of this problem, in the general setting,
has been
extensively developed (see [15]). Note, finally, that it is not necessary for a
spacetime $M$ to be time-orientable in order to `mimic' the above constructions
in a
sensible way. For non-time-orientable $M$ we still have a notion of
kinking{\footnote{Likewise, the existence of a globally non-singular Lorentz
metric on
a non-time-orientable spacetime is now equivalent to the existence of a global
non-vanishing {\underline{line field}} ${\{}v, -v{\}}$.}} (defined  now as the
degree
of the map from ${\Sigma}$ to {\underline{${\Bbb R}{\Bbb P}^{3}$}}) and
we still have a notion of `pinors', defined thus:}\\
\\
{\noindent {\bf Definition}~ {\it Let $M$ be a non-orientable spacetime, with
orthonormal frame bundle $O(M)$ a principal bundle with structure group $O(p,
q)$.
We say that $M$ has {\underline{pin-Lorentz structure}} iff there exists a
principal
bundle ${\bar O}(M)$ (with structure group Pin$(p, q)$) which is a $2 - 1$
covering of $O(M)$, so that the following diagram commutes:}}\\
\[
\begin{array}{cccccc}
{\mbox{Pin}}(p, q) &{\longrightarrow} &{\bar O}(M) &{\longrightarrow} &M & \\
{\downarrow} & &{\downarrow} & &{\downarrow} &{\mbox{\it identity}} \\
O(p, q) &{\longrightarrow} &O(M) &{\longrightarrow} &M &
\end{array}
\]
\\
{\noindent {\it where either (i) $p = 1, q = 3$, or (ii) $p = 3, q = 1$.}
\\
As we shall see, the topological obstruction to pin-Lorentz structure is
related to
that of spin-Lorentz structure.}\\
\vspace*{0.6cm}

\begin{center}
{\bf II. ~Equivalence of tetrad and spin-Lorentz cobordism}\\
\end{center}

{One of the first questions that comes to mind is whether or not there is any
connection between tetrad and spin-Lorentz cobordism. That there should be some
relation is implied by a theorem of Geroch [6]. One simple approach to this
question
would be to calculate the topological obstruction to tetrad cobordism and
compare it
to the obstruction to spin-Lorentz cobordism. However, this is not necessary
because
we have the following}\\
\\
{\noindent {\bf Theorem} (Hirzebruch and Hopf, [9]) {\it Let $M$ be a smooth,
compact, orientable four-manifold. Then $M$ is parallelizable iff $p_{1}(M) =
0, ~e(M)
= 0$, and $w_{2}(M) = 0$, where $p_{1}(M)$ is the first Pontryagin number,
$e(M)$ is
the Euler number, and $w_{2}(M)$ is the second Stiefel-Whitney class.}}\\
\vspace*{0.2cm}

{Now, when $M$ has non-trivial boundary ${\partial}M ~{\cong}~ {\Sigma}_{1}
{}~{\cup}~
{\Sigma}_{2} ~{\cup}~ ... ~{\cup}~ {\Sigma}_{n} ~{\not=}~  {\emptyset}, ~w_{2}$
$(M)$ is
defined as usual, but the relationship betwen $e(M)$ and the zeros of smooth
vector
fields on $M$ changes [4]. That is, if $v$ is a smooth vector field on $M$, and
${\sum} ~i_{v}$ denotes `the sum of the indices of $v$', then $e(M)$ is given
by}
\begin{equation}
{{\sum} ~i_{v} = e(M) ~+~ {\mbox{kink}}({\partial}M; v)}
\end{equation}

{Furthermore, for a manifold with non-empty boundary (of disjoint closed,
orientable
three-manifolds) we automatically have}
\[
{p_{1}(M) = 0 .}
\]

{Thus, amending the above theorem to deal with the case  when ${\partial}M
{}~{\not=}~
{\emptyset}$, we obtain}\\
\\
{\noindent {\bf Corollary}~ {\it Let $M$ be a smooth, compact orientable
four-manifold with non-empty boundary}}
\[
{{\partial}M ~{\cong}~ {\Sigma}_{1} ~{\cup}~ {\Sigma}_{2} ~{\cup}~ ... ~{\cup}~
{\Sigma}_{n} ~{\not=}~ {\emptyset} .}
\]
{{\it Then $M$ is parallelizable by a tetrad field {\{}$v_{i}:~ i = 1,
...4${\}} iff
$w_{2}(M) = 0$ and ${\sum} ~i_{v_{i}} = 0$, for any vector $v_{i}$ in the
tetrad.}}\\
\vspace*{0.2cm}

{Now, notice that if $v$ is a timelike vector field on $M$ with respect to a
Lorentz
metric $g_{ab}$, then the metric is globally non-singular iff}
\[
{{\sum} ~i_{v} = 0 .}
\]
{Furthermore, $M$ admits a spin structure iff $w_{2}(M) = 0$. Thus, the
obstructions
to $(M, g_{ab})$ being a spin-Lorentz manifold are precisely the obstructions
to $M$
being parallelizable, and so spin-Lorentz and tetrad cobordism are
equivalent.}\\
\vspace*{0.6cm}

\begin{center}
{\bf III. ~Derivation of the obstruction to spin-Lorentz cobordism}\\
\end{center}

{Since we are concerned with the obstruction to spin-Lorentz cobordism, it is
useful to
first review the second Stiefel-Whitney class $w_{2}(M)$ (the obstruction to
spin
structure on $M$).

Hence, suppose we are given a four-dimensional orientable manifold $M$ with
tangent
bundle $TM$. Given the $2 - 1$ covering map}
\[
{{\rho}: ~{\mbox{Spin}}(4) ~{\longrightarrow}~ SO(4)}
\]
{we can define, for transition function $h_{ab} ~{\in}~ SO(4)$, the lifting
${\bar
{h_{ab}}}  ~{\in}~ {\mbox{Spin}}(4)$, satisfying}
\[
{{\rho}({\bar {h_{ab}}}) = h_{ab}, ~~{\bar {h_{ab}}} = {\bar {h_{ab}}}^{-1}.}
\]
{By local triviality, such a lifting can always be found (locally).

Because ${\rho}$ is a homomorphism, and using the compatability condition, we
see that}
\[
{{\rho}({\bar {h_{ab}}} ~{\bar {h_{bc}}} ~{\bar {h_{ca}}}) = h_{ab} ~h_{bc}
{}~h_{ca} =
{\mbox{Id}},}
\]
{where Id is identity map on $U_{a} ~{\cap}~ U_{b} ~{\cap}~ U_{c} ~{\subset}~
M$
($U_{i}$s are open sets). Thus, ${\bar {h_{ab}}} ~{\bar {h_{bc}}} ~{\bar
{h_{ca}}}$ is
in the kernel of ${\rho}_{j}$; however, there is a sign ambiguity in the
kernel:}
\begin{equation}
{{\ker} = {\{ ~{\pm} {\mbox{Id}}~ \}}.}
\end{equation}
{However, for the ${\bar {h_{ab}}}$'s to define a global spin bundle over $M$,
they
must also satisfy the compatibility condition}
\begin{equation}
{{\bar {h_{ab}}} ~{\bar {h_{bc}}} ~{\bar {h_{ca}}} = {\mbox{Id}}}
\end{equation}

{Hence, define the ${\breve{\rm C}}$ech 2-cochain}
\[
{w_{2}(M) ~{\equiv}~ w_{2}(M; U_{i}, U_{j}, U_{k}): ~U_{i} ~{\cap}~ U_{j}
{}~{\cap}~
U_{k} ~{\longrightarrow}~ {\Bbb Z_{2}}}
\]
{via the relation}
\begin{equation}
{{\bar {h_{ab}}} ~{\bar {h_{bc}}} ~{\bar {h_{ca}}} = w_{2}(M)~{\mbox{Id}}}
\end{equation}
{(where ${\Bbb Z_{2}}$ here is multiplicative). Then $w_{2}(M) ~{\in}~ H^{2}(M;
{\Bbb Z_{2}})$ is called the {\it second Stiefel-Whitney class}.

Taking ${\Bbb Z_{2}}$ to be additive we have the easy}\\
\\
{\noindent {\bf Lemma 1.}~ {\it Let $M$ be as above. Then there exists a spin
bundle over $M$ iff}}
\[
{w_{2}(M) = 0.}
\]
\vspace*{0.1cm}

{Clearly, if $M$ admits Spin(4) spin structure and $M$ is Lorentz, then $M$
admits
$SL(2, {\Bbb C})$ spin structure and is a spin-Lorentz manifold.

To see how we can obtain a topological obstruction to spin-Lorentz structure on
$M$
which depends only on boundary data defined on ${\partial}M$, recall the
following}\\
\\
{\noindent {\bf Lemma} ~(Milnor and Kervaire, [5], page 517) ~{\it Let $M$ be
an
orientable, smooth manifold of dimension 4. Let $u({\partial}M)$ (the mod 2
Kervaire
semicharacteristic) be given by}}
\[
{u({\partial}M) = {\dim}_{{\Bbb Z_{2}}}(H_{0}({\partial}M; {\Bbb Z_{2}})
{}~{\oplus}~
H_{1}({\partial}M; {\Bbb Z_{2}})) {\mbox{\it mod}} ~2}
\]
{{\it Then the rank of the intersection pairing $h:~ H_{2}(M; {\Bbb Z_{2}})
{}~{\times}~
H_{2}(M; {\Bbb Z_{2}}) ~{\longrightarrow}~ {\Bbb Z_{2}}$ satisfies
${\mbox{rank}}(h) =
(u({\partial}M) ~+~ e(M)) \bmod 2$.}}\\
\vspace*{0.2cm}

{To see how the Lemma relates to spin-Lorentz structure, recall [12]
that the rank of the intersection pairing also satisfies}
\begin{equation}
{({\mbox{rank}}(h)) ~{\bmod 2} = 0 ~{\Longleftrightarrow}~ w_{2}(M) = 0}
\end{equation}
{Combining equation (10) with the above Lemma, we obtain}\\
\\
{\noindent {\bf Theorem 1.} {\it Let ${\Sigma}_{1}, {\Sigma}_{2}, ...
{\Sigma}_{n}$ be a collection of closed, orientable three-manifolds. Then there
exists
a spin-Lorentz cobordism, $M$, for ${\Sigma}_{1}, {\Sigma}_{2}, ...
{\Sigma}_{n}$ if
and only if}}
\begin{equation}
{(u({\partial}M) ~+~ {\mbox{\it kink}}({\partial}M; v)) ~{\mbox{\it mod}}~ 2 =
0 ,}
\end{equation}
{{\it where $u({\partial}M)$ is as above, and $v$ is the timelike vector
determined by the
metric on $M$.}}\\
\\
{\noindent {\it Proof.}~ Suppose such a spin-Lorentz cobordism, $M$,
exists. Then $M$ admits spin structure and so $w_{2}(M) = 0$; by equation (10),
$({\mbox{rank}}(h)) ~{\bmod ~2} = 0$. Hence $(u({\partial}M) ~+~ e(M)) ~{\bmod
{}~2} = 0$,
by the Lemma.

Furthermore, since $M$ is a Lorentz manifold with timelike  vector $v$ we must
have
${\sum} ~i_{v} = 0$ (since $v$ must not vanish); hence, by equation (6)}
\[
{-e(M) = {\mbox{kink}}({\partial}M; v)}
\]
{and so, modulo 2, we obtain}
\[
{(u({\partial}M) ~+~ {\mbox{kink}}({\partial}M; v)) ~{\bmod ~2} = 0 .}
\]

{Conversely, suppose that no such spin-Lorentz cobordism exists. Then any
cobordism
$M$ is one of three things: spin but not Lorentz, Lorentz but not spin, or
neither
spin nor Lorentz.

If $M$ is spin but not Lorentz, then $w_{2}(M) = {\mbox{rank}}(h) ~{\bmod ~2} =
0$  and
${\sum} ~i_{v} ~{\not=}~ 0$. If ${\sum} ~i_{v}$ is even, then we could take the
connected
sum of $M$ with a finite number of spin manifolds (of even Euler number) to
obtain a
spin cobordism $M^{\prime}$ with ${\sum} ~i_{v} = 0$ [1]. However, such an
$M^{\prime}$
would be a spin-Lorentz cobordism, contradicting our assumption. Thus, ${\sum}
{}~i_{v}$
must be odd and so we get}
\[
{(u({\partial}M) ~+~ {\mbox{kink}}({\partial}M; v)) ~{\bmod ~2} = 1}
\]

{Likewise, if $M$ is Lorentz but not spin, then ${\sum} ~i_{v} = 0$ and
$w_{2}(M) =
({\mbox{rank}}(h)) ~{\bmod ~2} = 1$ and so}
\[
{(u({\partial}M) ~+~ {\mbox{kink}}({\partial}M; v)) ~{\bmod ~2} = 1 .}
\]

{Finally, if $M$ is neither spin nor Lorentz, then ${\sum} ~i_{v} ~{\not=}~ 0$
and
$w_{2}(M) = ({\mbox{rank}}(h)) ~{\bmod ~2} = 1$. If ${\sum} ~i_{v}$ is even,
then
$(u({\partial}M) ~+~ {\mbox{kink}}({\partial}M; v)) ~{\bmod ~2} = 1$. Thus,
suppose
that ${\sum} ~i_{v}$ is odd. Recall that although $M$ is not spin, we can
always find
[8] a spherical modification of $M, M^{\prime}$, which is spin (a spin
cobordism always
exists). However, such a cobordism would satisfy $e(M^{\prime}) ~{\not=}~ e(M)
{}~{\bmod
{}~2}$. Thus, if $v^{\prime}$ is the vector field  $v$ extended to
$M^{\prime}$, we have
that ${\sum} ~i_{v^{\prime}}$ is even. However, $M^{\prime}$ is then a spin
manifold
with ${\sum} i_{v^{\prime}}$ an even number. This case was dealt with above,
and we
saw that}
\[
{(u({\partial}M) ~+~ {\mbox{kink}}({\partial}M; v)) ~{\bmod ~2} = 1 .}
\]

{This exhausts all possibilities. \hfill ${\square}$}\\
\\
{\indent Using Section II we obtain}\\
\\
{\noindent {\bf Corollary 1.} {\it Let ${\Sigma}_{1}, {\Sigma}_{2}, ...
{\Sigma}_{n}$ be a collection of closed, orientable three-manifolds. Then there
exists
a tetrad cobordism $M$, with global tetrad field {\{}$v_{1}, v_{2}, v_{3},
v_{4}${\}},
if and only if}}
\[
{(u({\partial}M) ~+~ {\mbox{\it kink}}({\partial}M; v_{i})) ~{\mbox{\it mod}}
{}~2 = 0}
\]
{{\it for any $v_{i}$ in the tetrad field.}}\\
\vspace*{0.2cm}

{Thus, we see that spin-Lorentz and tetrad cobordisms between arbitrary
three-manifolds
always exist, as long as we allow for arbitrary kink number (boundary data).}\\
\vspace*{0.6cm}

\begin{center}
{\bf IV. ~Discussion of Clifford algebras}\\
\end{center}

{Before discussing pin structures on non-orientable spacetimes, it is useful to
review
the Clifford algebras which give rise to the `Cliffordian Pin
groups'{\footnote{Note:
There are other $2 - 1$ covers of $O(p, q)$ which do not arise from any
Clifford
algebra. These give us `non-Cliffordian' pin structures. The obstructions to
non-Cliffordian pin structures have been worked out elsewhere [16]. See [20]
for an
excellent discussion of the different pin groups.}}, and to discuss some of the
subtleties associated with these groups.

Thus, let $(M, g_{ab})$ be any spacetime (not necessarily orientable). Then the
tangent bundle of $M, {\tau}_{M}$, can always be reduced to a bundle with
structure
group $O(3, 1)$ (for signature $(- + + +)$) or $O(1, 3)$ (for signature $(+ - -
-)$)
(actually $O(3, 1) ~{\simeq}~ O(1, 3)$, but as we shall see it is important
that we
keep the distinction between the signatures when we pass to the double covers
of these
groups). Now, we associate to the tangent space of $(M, g_{ab})$, at some $p
{}~{\in}~
M$, the `Clifford algebra', $Cl(T_{p}(M), g_{ab})$, which can be described as
follows:

Let ${\{}e_{1}, e_{2}, e_{3}, e_{4}{\}}$ be an orthonormal basis (with respect
to
$g_{ab}$) for $T_{p}(M)$. Then $Cl(T_{p}(M), g_{ab})$ is the algebra generated
by
${\{}e_{i}|i = 1, ... 4{\}}$, subject to the following relation:}
\[
{e_{i}e_{j} ~+~ e_{j}e_{i} = 2g(e_{i}, e_{j})}
\]

{Now, associated to any Clifford algebra $Cl(p, q)$ is the group of invertible
elements, $Cl_{*}(p, q)$. Let $P(p, q) ~{\subseteq}~ Cl_{*}(p, q)$ be
the subgroup generated by non-null vectors $v ~{\in}~ T_{p}(M)$ (i.e.,
$g_{ab}v^{a}v^{b}
{}~{\not=}~ 0$). Then ${\mbox{Pin}}(p, q) ~{\subseteq}~ P(p, q)$ is the
subgroup generated by elements $v ~{\in}~ T_{p}(M)$ with $g_{ab}v^{a}v^{b} =
{\pm}1$.
Thus, any element $x ~{\in}~ {\mbox{Pin}}(p, q)$ can be written as some
product:
$x = v_{1}v_{2} ... v_{n}$, where all the $v_{i}$'s are unit spacelike or
timelike
vectors. But we know that the groups ${\mbox{Pin}}(p, q)$ are double covers of
the groups $O(p, q)$, and so in some sense the pin groups must `re-express' all
of the
information contained in the Lorentz group (in fact, they must `re-express' the
information in a `simply connected' way, since ${\pi}_{1}(O(p, q)) ~{\simeq}~
{\Bbb Z_{2}}$
and ${\pi}_{1}({\mbox{Pin}}(p, q)) ~{\simeq}~ 0$). In fact, we see how elements
of the pin groups represent Lorentz transformations when we recall the
following:}\\
\\
{\noindent {\it Fact.}~ Any element of $O(p, q)$ can be represented as a
product of reflections across a finite number of (non-null) planes through the
origin
$O ~{\in}~ T_{p}(M)$.

Thus, let $x = v_{1}v_{2} ... v_{n}$ be any element of ${\mbox{Pin}}(p, q)$.
For
each vector $v_{i}$, let $v_{i}^{\perp}$ denote the plane perpendicular to
$v_{i}$.
Then, for any element $w ~{\in}~ T_{p}(M)$, the reflection of $w$ about
$v_{i}^{\perp}$ is given as}
\[
{w ~{\longrightarrow}~ w - 2(w ~{\bf {\cdot}}~ v_{i})v_{i}}
\]
{Hence, we can view $v_{1}v_{2} ... v_{n} ~{\in}~ {\mbox{Pin}}(p, q)$ as a
series of reflections about the planes $v_{n}^{\perp}, v_{n - 1}^{\perp}, ...
v_{1}^{\perp}$, i.e., $v_{1}v_{2} ... v_{n}$ has a natural interpretation as a
Lorentz
transformation.

Furthermore, we see that to every Lorentz transformation there correspond two
distinct
elements of ${\mbox{Pin}}(p, q)$. For example, if $T ~{\in}~ O(p, q)$
represents
time reversal, and $e_{1}$ is the (basis) unit timelike vector, then both
$e_{1}$ and
$-e_{1}$ correspond to $T$. And so on.

In the next section, we shall concentrate on the cobordism problem for
Cliffordian
pin bundles, i.e., bundles whose structure group can be obtained from a
Clifford
algebra (in the way described above). We note, however, that the cobordism
problem
for {\it non}-Cliffordian pin structures has been worked out elsewhere ([16]).

Indeed, these results can perhaps be taken as further evidence that there is no
immediate reason why we should insist that our underlying spacetime manifold be
orientable; we can do fermionic physics on non-orientable spacetimes using pin
bundles (see [17], [18], and in particular [19]). This point is especially
relevant
if we take the `spacetime foam' picture seriously, since there is no {\it a
priori}
reason why nature should prefer orientable fluctuations over non-orientable
fluctuations.}\\
\vspace*{0.6cm}

\begin{center}
{\bf V. ~Derivation of the obstructions to Cliffordian}\\
{\bf pin-Lorentz cobordism}\\
\end{center}

{In order to apply the above Lemma of Milnor and Kervaire to the derivation of
obstructions to pin-Lorentz structure, we first must derive some identities for
the
Stiefel-Whitney classes $w_{1}(M)$ and $w_{2}(M)$ when $M$ is non-orientable,
i.e.,
when $w_{1}(M) = 1$.

First, recall Wu's formula}
\begin{equation}
{w_{k}(M) = {\displaystyle{\sum_{i + j = k}}} Sq^{i}(v_{j})}
\end{equation}
{where `$Sq^{i}$' is the Steenrod squaring operation [10], and $v_{j} ~{\in}~
H^{j}(M)$ is the unique element which satisfies}
\begin{equation}
{(v_{j} ~{\smile}~ x)[w] = Sq^{j}(x)[w], ~{\forall} ~x ~{\in}~ H^{n - j}(M)}
\end{equation}
{where `${\smile}$' denotes cup product, and $w ~{\in}~ H_{4}(M)$ is the
fundamental
homology class [10]. Using equations (12) and (13), together with the Axioms
for
Steenrod squaring [10], we obtain:}
\begin{equation}
{w_{1}(M) = v_{1}}
\end{equation}
\begin{eqnarray}
w_{2}(M) &=& Sq^{0}(v_{2}) ~+~ Sq^{1}(v_{1}) ~+~ Sq^{2}(v_{0})\nonumber \\
          &=& v_{2} ~+~ v_{1} ~{\smile}~ v_{1} \\
          &=& v_{2} ~+~ w_{1} ~{\smile}~ w_{1}\nonumber
\end{eqnarray}
{Thus, let $x_{2} ~{\in}~ H^{2}(M; {\Bbb Z_{2}})$ be any 2-cochain, then}
\[
{w_{2}(M) ~{\smile}~ x_{2} = v_{2} ~{\smile}~ x_{2} ~+~ (w_{1} ~{\smile}~
w_{1})
{}~{\smile}~ x_{2}}
\]
{and since $v_{2} ~{\smile}~ x_{2} = x_{2} ~{\smile}~  x_{2}$, we get}
\begin{equation}
{w_{2}(M) ~{\smile}~ x_{2} = x_{2} ~{\smile}~ x_{2} ~+~ (w_{1} ~{\smile}~
w_{1})
{}~{\smile}~ x_{2}}
\end{equation}

{Now, recall the definition of `intersection pairing' between two cycles $x, y
{}~{\in}~
H_{2}(M; {\Bbb Z_{2}})$. First, let $x_{2}$ and $y_{2}$ be the 2-cochains
associated with $x$
and $y$, defined via}
\begin{eqnarray}
x_{2} ~{\frown}~ w &=& x\nonumber \\
y_{2} ~{\frown}~ w &=& y
\end{eqnarray}
{where `${\frown}$' denotes cap product. Then we define the intersection
pairing, $h:~
H_{2}(M; {\Bbb Z_{2}}) ~{\times}~ H_{2}(M; {\Bbb Z_{2}}) ~{\longrightarrow}~
{\Bbb Z_{2}}$, via the
relation}
\[
{h(x, y) = x ~{\bf {\cdot}}~ y = (x_{2} ~{\smile}~ y_{2}) ~{\frown}~ w .}
\]
{Now, the question is: How does the parity of the rank of $h$ relate to
$w_{2}(M)$ and
$w_{1}(M)$? To see the answer, suppose ${\mbox{rank}}(h)$ was even. Then every
cycle
$x ~{\in}~ H_{2}(M)$ would have to have self intersection number zero, i.e.,}
\[
{x ~{\bf {\cdot}}~ x = 0 \hspace*{1cm} {\forall} ~x ~{\in}~ H_{2}(M; {\Bbb
Z_{2}})
{}~~~{\Longrightarrow}~}
\]
\[
{x ~{\bf {\cdot}}~ x = (x_{2} ~{\smile}~ x_{2}) ~{\frown}~ w = 0, ~{\forall}
{}~x_{2}
{}~{\in}~ H^{2}(M; {\Bbb Z_{2}}) ~~~{\Longrightarrow}}
\]
\[
{((w_{2}(M) ~{\smile}~ x_{2}) ~-~ [w_{1} ~{\smile}~ w_{1}] ~{\smile}~ x_{2})
{}~{\frown}~ w = 0 ~~~{\Longrightarrow}}
\]
\begin{equation}
{w_{2}(M) - w_{1} ~{\smile}~ w_{1}  = 0}
\end{equation}
{Conversely, if ${\mbox{rank}}(h)$ was odd, then ${\exists} ~x ~{\in}~ H_{2}(M;
{\Bbb Z_{2}})$
such that $x ~{\bf {\cdot}}~ x ~{\not=}~ 0$, and so}
\begin{equation}
{w_{2}(M) - w_{1} ~{\smile}~ w_{1} ~{\not=}~ 0}
\end{equation}
{Combining (18) and (19), we obtain}
\begin{equation}
{w_{2}(M) ~+~ w_{1} ~{\smile}~ w_{1} = 0 ~~{\Longleftrightarrow}~~
{\mbox{rank}}(h) = 0
{}~\bmod 2}
\end{equation}

{Combining (20) with the Lemma of Milnor and Kervaire (which still holds, since
everything is in ${\Bbb Z_{2}}$ coefficients), we get}
\begin{equation}
{(u({\partial}M) + {\mbox{kink}}({\partial}M; g_{ab})) ~{\bmod 2} = 0
{}~{\Longleftrightarrow}~ (w_{2}(M) + w_{1} ~{\smile}~ w_{1}) = 0}
\end{equation}
{where ${\mbox{kink}}({\partial}M; v)$ is, again, the degree of the total map
${\partial}M ~{\longrightarrow}~ {\Bbb R}{\Bbb P}^{3}$ (or $S^{3}$ if $M$ is
time-orientable) defined
by $v$.

We are now in position to derive our topological obstructions (which depend
only upon
boundary data, choice of orientation, choice of signature, and behaviour of
1-cocycles under the cup product) using results of M. Karoubi [11]. We
therefore begin
by dividing the possible cases according to signature.}\\
\\
{\noindent {\bf Case 1.} (Signature $(- +  +  +)$)}

{In the case when the signature is $(- + + +)$ we get the following result:}

{First recall that the tangent bundle ${\tau}_{M}$ decomposes into a direct sum
of
subbundles,}
\[
{{\tau}_{M} ~{\cong}~ {\tau}^{+} ~{\oplus}~ {\tau}^{-}}
\]
{where ${\tau}^{+}$ is the `spacelike subbundle' and ${\tau}^{-}$ is the
`timelike
subbundle' (the terms refer to the behaviour of sections of these bundles with
respect to the Lorentz metric $g_{ab}$).

{\noindent By elementary Axioms [10] we have}
\begin{equation}
{w_{1}({\tau}_{M}) = w_{1}({\tau}^{+}) ~+~ w_{1}({\tau}^{-})}
\end{equation}
\[
{w_{2}({\tau}_{M}) = w_{2}({\tau}^{+}) ~+~ w_{2}({\tau}^{-}) ~+~
w_{1}({\tau}^{+})
{}~{\smile}~w_{1}({\tau}^{-})}
\]
{We shall often use the abbreviations $w_{1}({\tau}^{+}) = w_{1}^{+},
w_{1}({\tau}^{-}) = w_{1}^{-}, w_{2}({\tau}^{+}) = w_{2}^{+}$, etc.

Now, it is a theorem of Karoubi [11] that there is ${\mbox{Pin}}(3, 1)$
structure on $M$ if and only if the following equation holds:}
\begin{equation}
{w_{2}^{-} ~+~ w_{2}^{+} ~+~ w_{1}^{-} ~{\smile}~ w_{1}^{-} ~+~ w_{1}^{-}
{}~{\smile}~
w_{1}^{+} = 0}
\end{equation}
{Combining equations (22) and (23), we thus see that $M$ has ${\mbox{Pin}}(3,
1)$
structure if and only if}
\begin{equation}
{w_{2}(M) = w_{2}({\tau}_{M}) = w_{1}^{-} ~{\smile}~ w_{1}^{-}}
\end{equation}

{Combining equations (21), (22), and (24) we then have}\\
\\
{\noindent {\bf Theorem 2.} {\it Let ${\Sigma}_{1}, {\Sigma}_{2}, ...
{\Sigma}_{n}$ be a collection of closed three-manifolds. Then there exists a
pin-Lorentz cobordism, $M$ (of signature $(- + + +)$), for {\{}${\Sigma}_{i}:~
i = 1,
... n${\}} if and only if the following holds:}}
\[
{(u({\partial}M) + {\mbox{\it kink}}({\partial}M; g_{ab})) ~{\mbox{\it mod}} ~2
= 0
{}~~{\Longleftrightarrow}~~ w_{1}^{+} ~{\smile}~ w_{1}^{+} = 0}
\]
\\
{\noindent {\it Proof.}~ Suppose such a pin-Lorentz cobordism, $M$, exists.
Then equation (24) holds. Combining this with equation (21) we get}
\[
{(u({\partial}M) ~+~ {\mbox{kink}}({\partial}M; g_{ab})) ~{\bmod ~2} = 0
{}~~{\Longleftrightarrow}~~ w_{1}^{+} ~{\smile}~ w_{1}^{+} = 0}
\]

{The converse is also immediate. \hfill ${\square}$}
\vspace*{0.2cm}

{\indent We interpret Theorem 2 in the following subcases and examples (all of
which
deal with signature $(- + + +)$.}\\

{\noindent 1. First of all, suppose that we want our cobordism to be both space
and
time-orientable. Then we must have $w_{1}^{+} = 0$ and $w_{1}^{-} = 0$, and so
$w_{1}^{+} ~{\smile}~ w_{1}^{+} = 0$.  Thus, such a cobordism exists if and
only if}
\[
{(u({\partial}M) ~+~ {\mbox{kink}}({\partial}M; g_{ab})) ~{\bmod ~2} = 0.}
\]
\\
{\noindent 2. Similarly, if we insist that our cobordism be neither space nor
time-orientable, then we have $w_{1}^{+} = 1 = w_{1}^{-}$ and so $w_{1}^{+}
{}~{\smile}~
w_{1}^{+} = 1$. Thus, such a cobordism exists if and only if}
\[
{(u({\partial}M) ~+~ {\mbox{kink}}({\partial}M; g_{ab})) ~{\bmod ~2} = 1}
\]
\\
{\noindent 3. Now, however, suppose that we want our cobordism to be
time-orientable
but not space-orientable, i.e., $w_{1}^{-} = 0$ but $w_{1}^{+} = 1$. Then we
run into
various subtleties which are caused by the definition of the cup product,
${\smile}$.
To understand this, let us recall how the cup product is defined simplicially:

First, suppose that $a_{1}, b_{1}$ are two 1-cochains; then their cup product
is a
2-cochain which may be defined by its action on a singular simplex $S:~ T^{2}
{}~{\longrightarrow}~ M$. That is, $S$ is a map which imbeds the convex set
${\{}a_{1},
a_{2}, a_{3} ~{\in}~ {\Bbb R}$ ~$|a_{i} ~{\geq}~ 0, ~a_{1} ~+~ a_{2} ~+~ a_{3}
= 1{\}} = T^{2}
{}~{\subset}~ {\Bbb R}^{3}$, into $M$ (i.e., a tetrahedron is determined by the
origin plus
three linearly independent points $a_{1}, a_{2}$, and $a_{3}$ in ${\Bbb
R}^{3}$).

Next, let $f(a_{1}, a_{2}, a_{3}) = (a_{1}, a_{2}, 0)$ denote the `front 1-face
of
$T^{2}$' (i.e., $(a_{1}, a_{2}, 0)$ is the triangle formed by $0, a_{1}$, and
$a_{2}$)
and let  $b(a_{1}, a_{2}, a_{3}) = (0, a_{2}, a_{3})$ denote the `back 1-face
of
$T^{2}$' (i.e., $(0, a_{2}, a_{3})$ is the triangle formed by $0, a_{2}$, and
$a_{3}$). Then $S ~{\bf {\circ}}~ f$ is the imbedded front 1-face of $S(T^{2})$
and $S
{}~{\bf {\circ}}~ b$ is the imbedded back 1-face. Thus, it makes sense to
define the cup
product, $a_{1} ~{\smile}~ b_{1}$, of $a_{1}$ and $b_{1}$ by the identity}
\[
{a_{1} ~{\smile}~ b_{1} [S] = (a_{1}[S ~{\bf {\circ}}~ f]) ~{\bf {\cdot}}~
(b_{1}[S
{}~{\bf {\circ}}~ b]) ~{\in}~ {\Bbb Z_{2}},}
\]
{that is, we calculate the value of the 1-cochain $a_{1}$ on the 1-cycle $S
{}~{\bf {\circ}}~ f$, and we multiply it (in ${\Bbb Z_{2}}$) by the value that
the 1-cochain
$b_{1}$ gives on $S ~{\bf {\circ}}~ b$. This gives us a number in ${\Bbb
Z_{2}}$.

Now, the problem that arises is the following: It may be (in the above
described
setting) that any two-cycle, $c ~{\in}~ H_{2}(M; {\Bbb Z_{2}})$, satsfies the
following
property (property `[$P$]'):

No matter how we deform $c$ (via a continuous deformation), it is always the
case that
the `front 1-face', $c_{1}$, and the `back 1-face', $c_{1}^{\prime}$, satisfy}
\[
{w_{1}^{+} ~{\smile}~ w_{1}^{+} [c] = (w_{1}^{+} [c_{1}]) ~{\bf {\cdot}}~
(w_{1}^{+}
[c_{1}^{\prime}]) = 1 ~{\bf {\cdot}}~ 0 = 0}
\]
{That is, it may be that we can have a cobordism $M$ which is time-orientable
$(w_{1}^{-} = 0)$, is not space-orientable $(w_{1}^{+} = 1)$, and yet still
satisfies
$w_{1}^{+} ~{\smile}~ w_{1}^{+} = 0$ !

In fact, this situation {\it does} occur, as seen in the following}\\
\\
{\noindent {\it Example.}~ Let $K$ denote the two-dimensional Klein bottle,
and $T^{2}$ denote the two-dimensional torus. Then we can form a spacetime $M
{}~{\simeq}~ K ~{\times}~ T^{2}$. Remove a disk from $M$ to obtain a spacetime
$M^{\prime} = M - D^{4}, ~{\partial}M^{\prime} = S^{3}$. Then
${\mbox{kink}}({\partial}M^{\prime}; g_{ab}) = 1$. But $u({\partial}M^{\prime})
= 1$,
and so}
\[
{(u({\partial}M^{\prime}) ~+~ {\mbox{kink}}({\partial}M^{\prime}; g_{ab}))
{}~{\bmod ~2}
= 0 .}
\]
{Now, we can always choose the Lorentz metric on $M^{\prime}$ so that
$M^{\prime}$ is
time-orientable but not space-orientable (signature $(- + + +)$, i.e., the
non-space
orientability comes from the $`K'$ part of $M^{\prime}$), and so $M^{\prime}$
admits
pin-Lorentz structure if and only $w_{1}^{+} ~{\smile}~ w_{1}^{+} = 0$, i.e.,
iff
for any 2-cycle, $c$, $w_{1}^{+} ~{\smile}~ w_{1}^{+} [c] = 0$. However, $K$
itself (viewed as a smoothly embedded 2-manifold in $M^{\prime}$) is evidently
a
2-cycle (satisfying $w_{1}^{+} ~{\smile}~ w_{1}^{+} ~[K] = 0$), and so
$M^{\prime}$
admits (global) pin-Lorentz structure, even though $w_{1}^{+} = 1$.}\\
\\
{\indent Thus, we see that we can have {\it any} kink number in the case
$w_{1}^{+} = 1, w_{1}^{-} = 0$ depending on the topology of the cobordism, $M$.
This
is summed up in the following}\\
\\
{\noindent {\bf Corollary}~ {\it Let ${\Sigma}_{1}, {\Sigma}_{2}, ...
{\Sigma}_{n}$ be a collection of closed three-manifolds. Then there exists a
pin-Lorentz cobordism, $M$ (signature $(- + + +)$), with $w_{1}^{+}(M) = 1$ and
$w_{1}^{-}(M) = 0$ if and only if}}
\[
(u({\partial}M) ~+~ {\mbox{\it kink}}({\partial}M; g_{ab})) ~{\mbox{\it mod}}
{}~2 = \left\{
\begin{array}{cl}
0, & {\mbox{\it if any 2-cycle, $c$, satisfies}}\\
   & {\mbox{\it Property [$P$]}}\\
   & \\
1, & {\mbox{\it otherwise}}
\end{array}
\right.
\]
\\
{\noindent 4. Finally, suppose we insist that our cobordism, $M$, be
space-orientable
but not time-orientable. Then $w_{1}^{+} = 0$ and $w_{1}^{-} = 1$, and so
$w_{1}^{+}
{}~{\smile}~ w_{1}^{+} = 0$ regardless of whether or not there is a two-cycle
in $M$
satisfying Property [$P$]. Thus, such a cobordism exists if and only if}
\[
{(u({\partial}M) ~+~ {\mbox{kink}}({\partial}M; g_{ab})) ~{\bmod 2} = 0.}
\]

{This exhausts all of the possibilities for signature $(- + + +)$.}\\
\\
{\noindent {\bf Case 2.} (Signature $(+ - - -)$)~ In this case, we see [11]
that Theorem 2 is still `true'; that is, it is still the case that there exists
a
pin-Lorentz cobordism, $M$ (signature $(+ - - -)$), for ${\{}{\Sigma}_{i}:~ i =
1, ...
n{\}}$ if and only if the following holds:}
\[
{(u({\partial}M) ~+~ {\mbox{kink}}({\partial}M; g_{ab})) ~{\bmod 2} = 0
{}~~{\Longleftrightarrow}~~ w_{1}^{+} ~{\smile}~ w_{1}^{+} = 0}
\]

{The difference now is that $w_{1}^{+}$ refers to the {\it timelike}
orientation. Thus, we get the same results in
{\it Case 2} as we did in {\it Case 1}, only
with the values of $w_{1}^{-}$ and $w_{1}^{+}$ interchanged.

We conclude with}\\
\\
{\noindent {\it Example.}~ (Gibbons) ~${\frac{S^{3} ~{\times}~ [0,
1]}{{\Bbb Z_{2}}}}$}\\

{Here, slice de Sitter spacetime  $(+ - - -)$ with two spacelike slices at
times $t =
{\pm} 1$, and identify the resulting three-spheres antipodally. One then
obtains a
space, $M$, which is topologically ${\frac{S^{3} ~{\times}~ [0, 1]}{{\Bbb
Z_{2}}}}$, as
shown: Fig. 2}\\
\\
{\noindent Clearly, then $M$ is a space-orientable spacetime which is not
time-orientable and has ${\partial}M ~{\cong}~ S^{3}$ spacelike
(kink$({\partial}M;
g_{ab}) = 0$). Hence, $M$ has pin-Lorentz structure (since it has no 2-cycles
satisfying Property [$P$]) and so $M$ is the standard example of the `creation
of a
spacelike $S^{3}$ from nothing' spacetime (for signature $(+ - - -)$), which we
shall
encounter below.}\\
\vspace*{0.6cm}

\begin{center}
{\bf VI. Applications of the obstructions}\\
\end{center}

{We are interested in seeing what restrictions our invariants
place on the homotopy type of the metric in standard
spacetime examples which are frequently encountered.

Let us first consider spin-Lorentz structure. Then one of
the first examples that springs to mind is the `creation from
nothing universe', i.e., visually: Fig. 3}\\
\\
{\indent Here, $M$ is a compact spin-Lorentz manifold with
single boundary component ${\Sigma}$, which is to be
interpreted as a `three-surface of simultaneity' with respect
to some universal time indexed by a Morse function $f:~ M
{}~{\longrightarrow}~ {\Bbb R}$ (so that $f^{-1}(x) ~{\cong}~
{\Sigma}$, for some $x ~{\in}~ {\Bbb R}$).

If ${\Sigma} ~{\cong}~ S^{3}$, we have}
\[
{u({\partial}M) = {\dim}_{{\Bbb Z_{2}}}(H_{0}(S^{3}; {\Bbb Z_{2}})
{}~{\oplus}~ H_{1}(S^{3}; {\Bbb Z_{2}})) ~{\bmod ~2} = 1
{}~~{\Longrightarrow}}
\]
{in order to have}
\[
{(u({\partial}M) ~+~ {\mbox{kink}}({\Sigma}; g_{ab})) ~{\bmod
{}~2} = 0}
\]
{we must have}
\[
{{\mbox{kink}}({\Sigma}; g_{ab}) = 1 ~{\bmod ~2}}
\]

{Thus, if the topology of our perceived three-surface of
simultaneity ${\Sigma}$ is $S^{3}$, then the metric $g_{ab}$
must have non-trivial homotopy type with respect to
${\Sigma}$ (in particular, there must exist an
{\it odd} number of kink regions of $g_{ab}$ with
respect to ${\Sigma}$).

On the other hand, if ${\Sigma} ~{\cong}~ S^{1} ~{\times}~
S^{2}$, then}
\[
{u({\Sigma}) = 0,}
\]
{and so we can have a creation from nothing universe with
kink$({\Sigma}; g_{ab}) = 0$, as long as ${\Sigma} ~{\cong}~
S^{1} ~{\times}~ S^{2}$. If we live in an expanding universe,
with global spin-Lorentz structure, and our perceived
three-surface of simultaneity ${\Sigma}$ is everywhere
spacelike (no kinking), then we must have ${\Sigma} ~{\cong}~
S^{1} ~{\times}~ S^{2}, {\Bbb R}{\Bbb P}^{3}$, or some other three-manifold
satisfying}
\[
{u({\Sigma}) = 0.}
\]

{Another example arises when one considers the creation of a
single `time machine' [6], in the sense of Thorne and
co-workers. Explicitly, Thorne et al. speculate that an
`advanced' civilization may someday be able to create a
spacetime wormhole (with spacelike topology $S^{1} ~{\times}~
S^{2}$) by `pulling' such a wormhole out of the quantum foam.
Thus, assuming that the initial topology (spacelike) of the
universe is $S^{3}$, we are concerned with what our invariant
tells us about the homotopy type of $g_{ab}$ with respect to
$S^{1} ~{\times}~ S^{2}$ (the `final' topology). Writing}
\[ \left\{ \begin{array}{l}
{\Sigma}_{i} ~{\cong}~ S^{3}\\
{\Sigma}_{f} ~{\cong}~ S^{1} ~{\times}~ S^{2}
\end{array}
\right. \]
{we are concerned with the spin-Lorentz cobordism, $M$, for
${\Sigma}_{i}$ and ${\Sigma}_{f}$. Since}
\[
{u({\partial}M) = u({\Sigma}_{i} ~{\cup}~ {\Sigma}_{f}) = 1}
\]
{we see that we must have}
\[
{{\mbox{kink}}({\partial}M) = 1 ~{\bmod 2}.}
\]

{Thus, the metric must have non-trivial homotopy on the
wormhole ${\Sigma}_{f}$, assuming ${\Sigma}_{i}$ was
spacelike.

Now let us consider pin-Lorentz structure. Then one of the
first things we notice is that we can have a creation from
nothing universe, $M$, with ${\partial}M ~{\cong}~ S^{3}$ and
${\mbox{kink}}({\partial}M; g_{ab}) = 0$. In other words, if
the signature is $(- + + +)$ then there exist compact
pin-Lorentz manifolds, which are either (i) neither space nor
time-orientable, or are (ii) time-orientable but not
space-orientable, and which have a single spacelike boundary
component homeomorphic to $S^{3}$.

Likewise, if the signature is $(+ - - -)$ then there exist
compact pin-Lorentz manifolds, which are either (i) neither
space nor time-orientable, or are (ii) space-orientable but
not time-orientable, and which have a single boundary
component homeomorphic to $S^{3}$.

Finally, let us again consider the `time machine' situation
of Thorne and co-workers, i.e., ${\partial}M ~{\cong}~ S^{3}
{}~{\cup}~ (S^{1} ~{\times}~ S^{2})$. Then we see that we can
now have pin-Lorentz spacetimes for which both the initial
slice, ${\Sigma}_{i} ~{\cong}~ S^{3}$, and the final slice,
${\Sigma}_{f} ~{\cong}~ S^{1} ~{\times}~ S^{2}$ are
{\it spacelike}. This is outlined as follows:

If the signature is $(- + + +)$ then there exist compact
pin-Lorentz manifolds $M_{i}$, which are either (i) neither
space nor time-orientable, or are (ii) time-orientable but
not space-orientable, and which have everywhere spacelike
boundaries ${\partial}M_{i} ~{\cong}~ S^{3} ~{\cup}~ (S^{1}
{}~{\times}~ S^{2})$.

Following the above example, a similar statement holds for
the case with signature $(+ - - -)$.

We conclude with a discussion on kinking and causality.}\\
\vspace*{0.6cm}

\begin{center}
{\bf VII. Kinking and causality}\\
\end{center}

{Suppose $M$ is a compact spacetime with}
\[
{{\partial}M  ~{\cong}~ {\Sigma}_{1} ~{\cup}~ {\Sigma}_{2} ~{\cup}~ ...
{}~{\cup}~
{\Sigma}_{n} ~{\not\cong}~ ~{\emptyset}.}
\]

{Recently, there has been some suspicion that there may be a relation between
the
topology of ${\partial}M$, along with the value of kink$({\partial}M; g_{ab})$,
and
the existence of closed timelike curves (CTCs) in $M$. In particular, it was
conjectured [2] that if ${\partial}M ~{\cong}~ S^{3}$ and kink$({\partial}M;
g_{ab})
= 0$, then there must exist CTCs in $M$.

In recent work [7] with Roger Penrose, we managed to show the above conjecture
to be
false (by counter example); in fact we proved the more general}\\
\\
{\noindent {\bf Theorem 3.}~ {\it Let ${\Sigma}_{1}, {\Sigma}_{2}, ...
{\Sigma}_{n}$ be any collection of closed, orientable three-manifolds, $n
{}~{\in}~
{\Bbb Z}$ an
arbitrary integer. Then there exists a compact causal spacetime $M$ with
${\partial}M
{}~{\cong}~ {\Sigma}_{1} ~{\cup}~ {\Sigma}_{2} ~{\cup}~ ... ~{\cup}~
{\Sigma}_{n}$ and
kink$({\partial}M; v) = n$, where $v$ is a timelike vector field.}}\\
\\
{\noindent {\bf Theorem 4.}~ {\it If $M$ is compact and causality violating,
with ${\partial}M ~{\cong}~ {\Sigma}_{1} ~{\cup}~ {\Sigma}_{2} ~{\cup}~ ...$
$~{\cup}~
{\Sigma}_{n} ~{\not\cong}~ ~{\emptyset}$, then there exists a
continuous deformation of the metric on $M$ such that the new spacetime with
deformed
metric does not possess CTCs.}}\\
\\
{\noindent {\it Proof.}~(of Theorem 3)~ Let ${\Sigma}_{1}, {\Sigma}_{2},
... {\Sigma}_{n}$ be any collection of closed, orientable three-manifolds, $n
{}~{\in}~
{\Bbb Z}$ any integer. Then we can always find a Lorentz manifold $M$ (with
metric $g_{ab}$
and timelike vector $v$) such that ${\partial}M ~{\cong}~ {\Sigma}_{1}$
$~{\cup}~
{\Sigma}_{2} ~{\cup}~ ... ~{\cup}~ {\Sigma}_{n}$ and
${\mbox{kink}}({\partial}M; v) =
n$. This follows from the general formula (6):}
\[
{e(M) = {\sum} ~i_{v} - {\mbox{kink}}({\partial}M; v)}
\]

{Now, we can cover $M$ with a finite number of sets $B_{p_{i}}$ of the form}
\[
{B_{p_{i}} = {\{}x ~{\in}~ I^{+}(p_{i}) ~{\cap}~ I^{-}(q)~|~q ~{\in}~
I^{+}(p_{i}){\}}}
\]

{Furthermore [13], we can take the sets in this finite cover to be fine enough
that
they are all locally causal (i.e., no CTC lies entirely in any one of the
$B_{p_{i}}$s).

Now, the crucial idea of the construction depends upon our ability to cut all
of the
CTCs by removing a finite number of four-balls. That we can do this is
reasonably
intuitively obvious, but we justify this construction more rigorously as
follows.

Begin by successively removing the `$t = 0$' Cauchy surface, $C_{i}$ from each
of our
locally causal covering sets $B_{p_{i}}$, as shown: Fig. 4}\\
\\
{\indent Now, at each stage $C_{i}$ may already be intersected by a previously
removed part
(assumed to be a union of three-disks), $R_{i - 1}$, so subdivide to get a
covering of
what is left by three-disks, ($D^{3}$s), as shown: Fig. 5}\\
\\
{\indent Next, modify $C_{i}$ according to the following two rules: Fig. 6}\\
\\
{\indent Adjoin the result to $R_{i - 1}$ to get $R_{i}$, which is thus given
as a disjoint
union of three-balls, $D_{j}^{3}$, as shown: Fig. 7}\\
\\
{\indent Finally, thicken out the $D_{j}^{3}$s to get disjoint four-balls
$B_{j}^{4}$s which
clearly cut the CTCs.

Hence, we can cut all of the CTCs with a finite number of such four-balls.

We now connect each of these little removed four-balls to the `old' boundary of
$M,
{\partial}M$, via little tubes $T_{j} ~{\cong}~ D^{3} ~{\times}~ [0, 1]$; that
is, we
cut out a little tube leading from some component of the `old' boundary of $M$
to the
new boundary component formed by removing a $B_{j}^{4}$, as shown: Fig. 8}\\
\\
{\indent Call the new manifold obtained after such a finite sequence of
operations `$N$'. Then
clearly}
\[
{{\partial}N ~{\cong}~ {\partial}M ~{\cong}~ {\Sigma}_{1} ~{\cup}~ {\Sigma}_{2}
{}~{\cup}~ ... ~{\cup}~ {\Sigma}_{n}}
\]
{since all we did to obtain $N$ was push a lot of `dimples' into the boundary
of $M$
(the boundary ${\partial}N$ is a continuous deformation of ${\partial}M$).

Furthermore, $v$ (and hence $g_{ab}$) is still global and non-vanishing on $N$,
i.e.,
${\sum} ~i_{v} = 0$. Thus, ${\mbox{kink}}({\partial}N; v) = n$. Thus, $N$ is a
causal
spacetime with ${\partial}N ~{\cong}~ {\Sigma}_{1} ~{\cup}~ {\Sigma}_{2}
{}~{\cup}~ ...
{}~{\cup}~ {\Sigma}_{n}$ and ${\mbox{kink}}({\partial}N; g_{ab}) = n$, and the
theorem
is proved.

\hfill ${\square}$}
\vspace*{0.2cm}

{\indent To prove Theorem 4, we continuously retract ${\partial}N$ back to
${\partial}M$ (via a homotopy) and `pull' the metric with the retraction (via
the
isotopy which lifts from the homotopy).

In closing, we note that Dr. R.P.A.C. Newman has strengthened the above proof
of
Theorem 3 by continuously retracting ${\partial}M$ all the way back to the
skeleton of
$M$. In this way, he is able to use fewer intuitive diagrams (and more
theorems) to
prove the result.}\\
\vspace*{0.6cm}

{\noindent \bf VIII. ~Conclusion}\\

{In closing, we point out some interesting questions which emerge from this
work.

First, as was shown in [1], the obstructions to spin-Lorentz and pin-Lorentz
structures can be interpreted physically as kinematical obstructions to the
creation
of certain types of `time machines'. Are there any other kinematical aspects of
physical law which one might also hope to apply to the question of the
theoretical
possibility of time travel (i.e., the Chronology Protection Conjecture)?

Second, does there exist any general relationship betwen kinking and geodesic
incompleteness? For example, a spherically symmetric (asymptotically flat) kink
region
has incomplete null geodesics corresponding to the roots of $g_{00}$ [3]. Can
one
develop a general statement which tells us `which' types of kinking with
respect to
`which' types of three-surfaces (in either compact or non-compact spacetimes)
inevitably lead to geodesic incompleteness?}\\
\vspace*{0.6cm}

{\noindent \bf Acknowledgements}\\

{The author would like to thank his advisor, Dr. G.W. Gibbons, Prof. S.W.
Hawking,
Prof. R. Penrose, Dr. R.P.A.C. Newman and Dr. L. Dabrowski for helpful
comments, ideas,
and criticisms during the compilation of this work. Also thanks to Jo Ashbourn
(Piglit)
for loving support and help with preparing this paper. This work was supported
by
NSF Graduate Fellowship No. RCD-9255644.}\\
\vspace*{0.6cm}

{\noindent \bf References}\\

{\noindent [1] G.W. Gibbons and S.W. Hawking, {\it Selection Rules for Topology
Change}, Commun. Math. Phys., {\bf 148}, No. 2, (1992).}\\
\\
{\noindent [2] G.W. Gibbons and S.W. Hawking, {\it Kinks and Topology Change},
Phys.
Rev. Lett., {\bf 69}, No. 12, (1992).}\\
\\
{\noindent [3] David Finkelstein and Gin McCollum, {\it Kinks and Extensions},
J.
Math. Phys., {\bf 16}, No. 11, (Nov. 1975).}\\
\\
{\noindent [4] V.I. Arnold, {\it Singularity Theory}, London Math. Soc. Lecture
Notes
No. {\bf 53}, CUP, (1981).}\\
\\
{\noindent [5] M.A. Kervaire and J.W. Milnor, {\it Groups of Homotopy Spheres:
I},
Annals of Math., {\bf 77}, No. 3, (May 1963).}\\
\\
{\noindent [6] R.P. Geroch, {\it Spinor Structure of Spacetimes in General
Relativity:
I}, J. Math. Phys., {\bf 9}, 1739--1744, (1968).}\\
\\
{\noindent [7] A. Chamblin and R. Penrose, {\it Kinking and Causality}, Twistor
Newsletter, No. {\bf 34}, 13--18, (May 1992).}\\
\\
{\noindent [8] J.W. Milnor, {\it On the Stiefel-Whitney numbers of complex
manifolds
and of spin manifolds}, Topology, {\bf 3}, 223--230, (1965).}\\
\\
{\noindent [9] G.S. Whiston, {\it Compact Spinor Spacetimes}, J. Phys. A: Math.
Gen.,
{\bf 11}, No. 7, (1978).}\\
\\
{\noindent [10] J.W. Milnor and J.D. Stasheff, {\it Characteristic Classes},
Princeton Univ.
Press, (1974).}\\
\\
{\noindent [11] M. Karoubi, {\it Algebres de Clifford et K-Theory}, Ann.
Scient. Ec.
Norm. Sup, $1^{e}$ serre, t.1, pg. 161, (1968).}\\
\\
{\noindent [12] H. Lawson and Marie-Louise Michelson, {\it Spin Geometry},
Princeton
Univ. Press, (1989).}\\
\\
{\noindent [13] R. Penrose, {\it Techniques of Differential Topology in General
Relativity}, Soc. for Industrial and Applied Mathematics, (1992).}\\
\\
{\noindent [14] N. Steenrod, {\it The Topology of Fibre Bundles}, Princeton
Univ.
Press, (1951).}\\
\\
{\noindent [15] R. Stong, {\it Notes on Cobordism Theory}, Princeton Math.
Notes,
Princeton Univ. Press, (1958).}\\
\\
{\noindent [16] A. Chamblin, {\it On the Obstructions to non-Cliffordian Pin
Structures}, DAMTP preprint No. R93/13, (1993).}\\
\\
{\noindent [17] S. Carlip and C. DeWitt-Morette, {\it Where the sign of the
metric
makes a difference}, Phys. Rev. Lett., {\bf 60}, pg. 1599, (1988).}\\
\\
{\noindent [18] C. DeWitt-Morette and Shang-Jr Gwo, {\it One Spin Group; Two
Pin
Groups}, Gauss Symposium, publ. by Gaussanium Inst., (1990).}\\
\\
{\noindent [19] C. DeWitt-Morette and B. DeWitt, {\it The Pin Groups in
Physics},
Phys. Rev. D, {\bf 41}, pg. 1901, (1990).}\\
\\
{\noindent [20] L. Dabrowski, {\it Group Actions on Spinors}, Monographs and
Textbooks
in Physical Science, Bibliopolis, (1988).}\\
\\

\end{document}